\documentclass[aps,prl,twocolumn,superscriptaddress,floatfix,amsmath,showpacs]{revtex4}

\usepackage{graphicx}
\usepackage{amssymb}

\begin{document}
\title{Bose-Einstein condensation of magnons under incoherent pumping}
\author{A.V.~Chumak}
 \email{chumak@physik.uni-kl.de}
 \altaffiliation[\\ Also at ]{Fachbereich Physik and Forschungszentrum
OPTIMAS, Technische Universit\"{a}t Kaiserslautern, 67663 Kaiserslautern, Germany}

\affiliation{Department of Radiophysics, National Taras Schevchenko University of Kiev, 01033 Kiev,
Ukraine}

\author{G.A.~Melkov}
\affiliation{Department of Radiophysics, National Taras Schevchenko University of Kiev, 01033 Kiev,
Ukraine}

\author{V.E.~Demidov}
\affiliation{Institute for Applied Physics, University of M\"{u}nster, 48149 M\"{u}nster, Germany}

\author{O.~Dzyapko}
\affiliation{Institute for Applied Physics, University of M\"{u}nster, 48149 M\"{u}nster, Germany}

\author{V.L.~Safonov}
\affiliation{Mag \& Bio Dynamics, Inc., Escondido, California, U.S.A.}

\author{S.O.~Demokritov}
\affiliation{Institute for Applied Physics, University of M\"{u}nster, 48149 M\"{u}nster, Germany}

\date{\today}

\begin{abstract}
Bose-Einstein condensation in a gas of magnons pumped by an incoherent pumping source is experimentally
studied at room temperature. We demonstrate that the condensation can be achieved in a gas of bosons
under conditions of incoherent pumping. Moreover, we show the critical transition point is almost
independent of the frequency spectrum of the pumping source and is solely determined by the density of
magnons. The electromagnetic power radiated by the magnon condensate was found to scale quadratically
with the pumping power, which is in accordance with the theory of Bose-Einstein condensation in magnon
gases.
\end{abstract}


\maketitle%
Several decades after the phenomenon of Bose-Einstein condensation (BEC) has been predicted for a gas of
atoms with an integer spin (bosons) \cite{ref1}, it was understood that bosonic elementary excitations
in solids, quantum liquids and giant molecules can undergo the same transition \cite{ref2, ref3, ref4}.
BEC takes place, if the density of (quasi)-particles is larger than a critical value, $N_\mathrm{c}$,
which increases with increasing temperature of the system. In this connection the advantage of bosons in
solids in comparison with real Bose atoms is that their density can be substantially increased by
external pumping and therefore the BEC transition can be reached at higher temperatures. Experimental
observation of BEC in ensembles of laser pumped excitons \cite{ref5} and polaritons \cite{ref6, ref7} as
well as in the gas of parametrically pumped magnons \cite{ref8} were reported at temperatures, which are
much higher than those for atomic gases. However, since bosonic quasi-particles in solids are pumped by
an external source, the source might introduce an additional coherence in the system. This effect can
mimic the spontaneous coherence intrinsic for BEC \cite{ref9}, in particular, if the pumping source is
coherent, as it was the case in the above experiments.

 In this Letter we investigate a gas of magnons pumped by an incoherent (noise) pumping source covering a relatively wide frequency interval. We report
the formation of BEC in the ensemble of magnons pumped by such a source, which is a principal step
showing that the coherent condensate of quasi-particles can be created by the magnons with
randomly-distributed frequencies, wave-vectors and phases. We experimentally show that the critical
point (which was documented by a vast increase (6-7 orders of magnitude) of the electromagnetic
radiation from the condensate) is almost independent of the frequency spectrum of the pumping source and
is solely determined by the density of magnons.

Possibility of BEC in a gas of magnons (the quanta of spin waves in magneto-ordered systems) under
strong microwave pumping was considered about twenty years ago \cite{ref10, ref11}. Recently BEC of
quasi-equilibrium magnons was discovered in yttrium iron garnet (YIG) thin films at room temperature
\cite{ref8}. The magnons in this and several following experiments \cite{ref12, ref13, ref14} were
pumped by a coherent microwave through parametric excitation: a microwave photon of the frequency
$f_\mathrm{p}$ creates a pair of primary magnons with the frequency, which is half of the pumping
frequency and oppositely oriented wave vectors
$f_\mathrm{\textbf{k}_p}=f_\mathrm{-\textbf{k}_p}=f_\mathrm{p}/2$. If the microwave power exceeds a
certain threshold, $P_1^0\propto(\delta f_\mathrm{p})^2$, where $\delta f_\mathrm{p}$ is the magnon
relaxation frequency, the number of parametric pairs grows dramatically in a narrow frequency interval
around $f_\mathrm{p}/2$. The primary magnons are ``very hot'': their spectral density becomes several
orders of magnitude larger than that of the thermal magnons at room temperature. Four-magnon scattering
processes, which preserve the total number of magnons, redistribute the magnons over the magnon spectrum
and create a quasi-equilibrium distribution of magnons outside of the resonance region.

\begin{figure}[t]
\begin{center}
\scalebox{1}{\includegraphics[width=7.5 cm,clip]{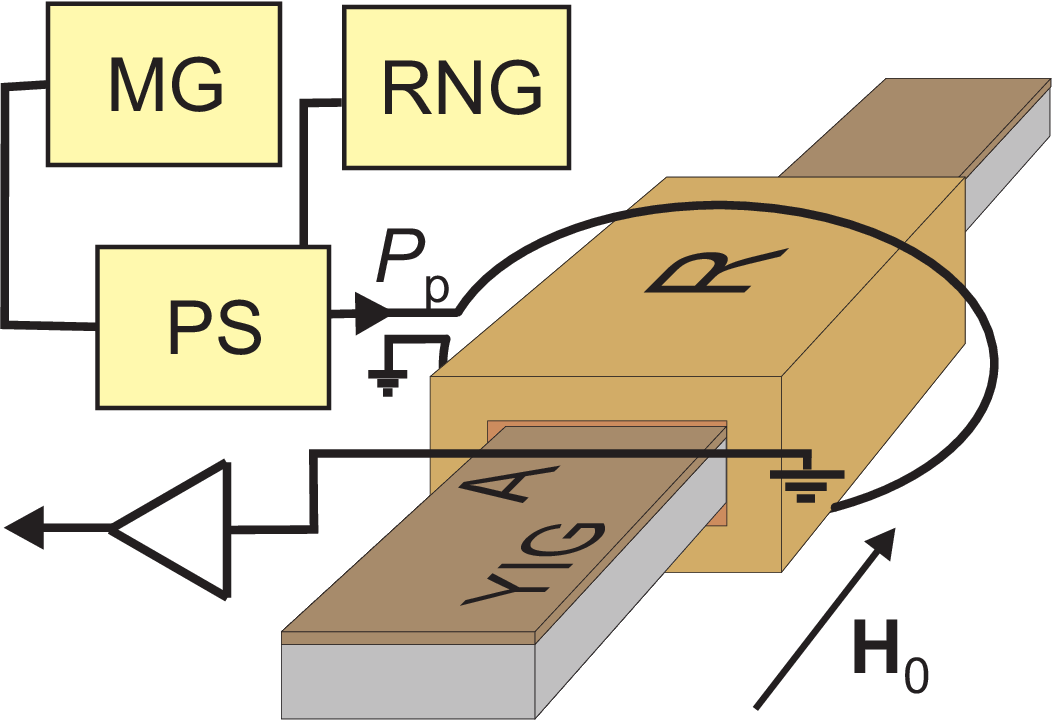}}
\end{center}
\vspace*{-0.4cm}\caption{(Color online) Schematic diagram of the used experimental setup. Microwave
generator (MG) generates pulses of microwave radiation. Phase-shifter (PS) controlled by the random
number generator (RNG) randomizes the phase of the radiation which is supplied to the dielectric
resonator (R). The resonator applies the noise pumping power to the YIG film. Antenna (A) detects the
electromagnetic radiation from the created condensate.} \label{fig1}
\end{figure}

The population function of magnons, as it follows from the theory \cite{ref11} and is confirmed by the
experiment \cite{ref12}, is the Bose-Einstein one with an effective temperature $T$ and an effective
chemical potential $\mu$. Without pumping $\mu=0$ and $T=T_\mathrm{lattice}$ (room temperature in our
case) due to thermal equilibrium between the magnon gas and the lattice. Applied pumping increases $\mu$
due to excitation of additional magnons. The value of $\mu$ is determined by the energy flow equilibrium
between the microwave pumping and the spin-lattice relaxation of magnons. Thus, by changing the pumping
power, one can control $\mu$. On the other hand, the deviation of effective temperature from the lattice
temperature can be neglected at room temperature as long as the total number of pumped magnons
($10^{18}-10^{19}\,\mathrm{cm}^{-3}$ in our experiments) is much smaller than the total number of
thermal magnons ($10^{22}\,\mathrm{cm}^{-3}$). An incoherent pumping covering the frequency interval
$f_\mathrm{p}\pm\Delta f/2$ does not modify the physical picture of the parametric pumping dramatically
\cite{ref11}, merely increasing the pumping power threshold $P_1\propto\delta f_\mathrm{p}(\delta
f_\mathrm{p}+\Delta f)$ \cite{ref15, ref16, ref17}.

When $\mu$ reaches the bottom of the spin-wave spectrum $\mu_\mathrm{c}=2\pi\hbar f_\mathrm{min}$, where
$f_\mathrm{min}=\mathrm{min}(f_\mathrm{\textbf{k}})$, the magnons at the lowest magnon state create a
coherent Bose-Einstein condensate. It can be revealed either by enormous light scattering cross-section
\cite{ref13} or by direct detection of the electromagnetic radiation from the condensate
 \cite{ref14}.

The sketch of the experimental setup is shown in Fig.~\ref{fig1}. The measurements were carried out on
$20\,\mu$m thick epitaxial yttrium-iron-garnet (YIG) film with the lateral dimensions
$1.5\times10\,\mathrm{mm^2}$. A uniform in-plane static magnetic field $H_0=0.8$\,kOe was applied along
the long side of the sample. The corresponding $f_\mathrm{min}=2.1$\,GHz. Magnons were excited by the
microwave field of an open dielectric resonator with resonance frequency $f_\mathrm{p}=9.4$\,GHz. The
microwave magnetic field was parallel to the static field to fulfill the conditions of the parallel
pumping. The resonator was fed with microwave pulses with $40\,\mu$s duration and a repetition period of
20\,ms. The incoherency of the pumping signal was realized by a random variation of its phase by
$\pm\pi/2$ by means of a phase shifter, controlled by a random number generator. The phase was
randomized with a frequency $f_\mathrm{d}$, which determines the width of frequency spectrum of the
pumping power $\Delta f\simeq0.9f_\mathrm{d}$. The width $\Delta f$ was typically much greater than the
relaxation frequency $\delta f_\mathrm{p}$ ($\sim1-2$\,MHz in our experiment) determined by the magnon
lifetime. The frequency-integrated peak pumping power $P_\mathrm{p}$ was about 5\,W.

To detect the electromagnetic radiation from the excited system, a microwave antenna of a width
$w=50\,\mu$m was attached to the surface of the film as shown in Fig.~\ref{fig1}. The microwave signal
received by the antenna was filtered within a window of 3.9-4.5\,GHz and was amplified by a low-noise
microwave amplifier.

\begin{figure}[b]
\begin{center}
\scalebox{1}{\includegraphics[width=7.5 cm,clip]{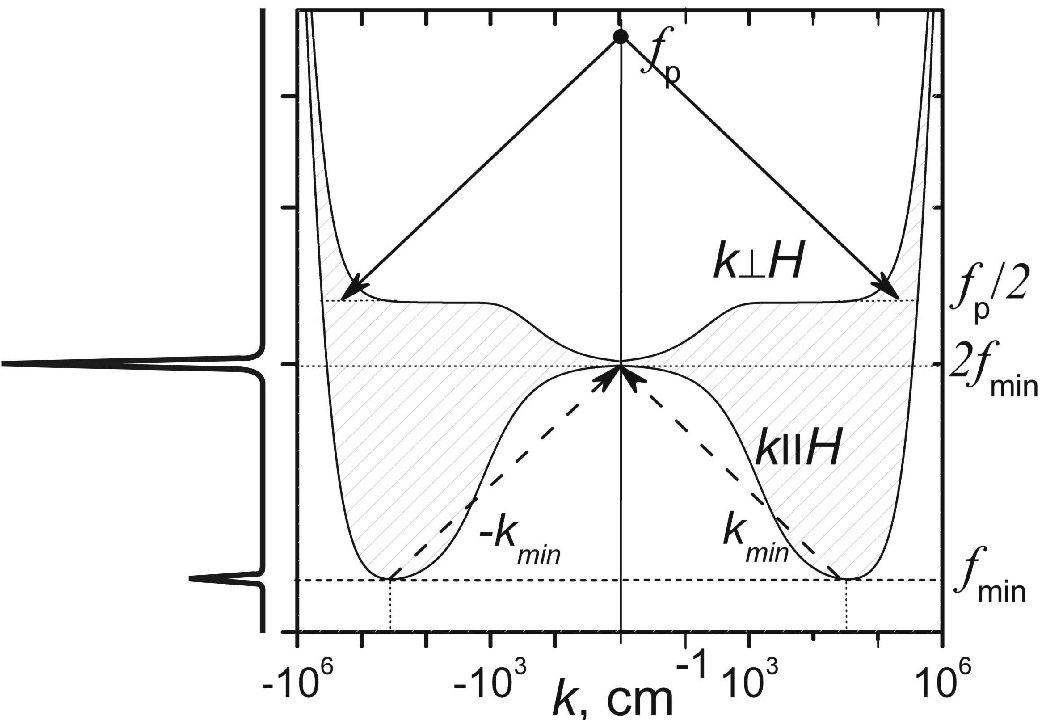}}
\end{center}
\vspace*{-0.4cm}\caption{Processes of parametric pumping used for magnon excitation and of magnon
confluence used for the detection of BEC. Right: magnon spectrum for $20\,\mu$m thick, magnetized
in-plane YIG film ($H_0=800$\,Oe). Solid arrows illustrate the parametric pumping process, dashed arrows
indicate the confluence of the magnons from the bottom of the spectrum, used for detection of BEC. Left:
sketch of the spectrum of the signal detected by the antenna.} \label{fig2}
\end{figure}

The condition $\mu(P_\mathrm{p})=\mu_\mathrm{c}=2\pi\hbar f_\mathrm{min}$ corresponds to another
threshold value of the pumping power $P_\mathrm{p}=P_2$. Above this value an essential part of magnons
condensates close to the bottom of the magnon spectrum, $f_\mathrm{min}$. This point of the magnon
spectrum, corresponding to the so-called backward volume magnetostatic waves, i.e.,
$\mathrm{\textbf{k}||\textbf{H}_0}$, is doubly degenerated (note $\textbf{k}_\mathrm{min}$ and
$-\textbf{k}_\mathrm{min}$ in Fig.~\ref{fig2}) with $k_\mathrm{min}\cong3\times10^4\,\mathrm{cm}^{-1}$.
We experimentally determine the critical power $P_2$ by detecting the electromagnetic radiation from the
bottom of the magnon spectrum by the antenna. Two spectral lines can be observed in such a case. First
one at $f_\mathrm{min}$ corresponds to the direct interaction of the magnon condensate with antenna.
This line is rather weak, since the antenna is much wider than the wavelength of the corresponding spin
waves $w \cdot k_\mathrm{min} \gg 1$ and different spatial parts of the condensate having different
phases compensate each other. Second line at $2f_\mathrm{min}$ is mainly created by the process of
two-magnon confluence ($\textbf{k}_\mathrm{min}$ and $-\textbf{k}_\mathrm{min}$) into uniform precession
of magnetization ($\textbf{k}=0$) as indicated in Fig.~\ref{fig2} by the dash arrows. This radiation can
be efficiently detected by the antenna since the phase of the precession with the double frequency is
almost uniform over the antenna. This spectral line was used in our experiment to detect the formation
of BEC. The sensitivity of the developed detection scheme in the frequency interval 3.9-4.5\,GHz was
about $10^{-14}$\,W.

\begin{figure}[t]
\begin{center}
\scalebox{1}{\includegraphics[width=7.5 cm,clip]{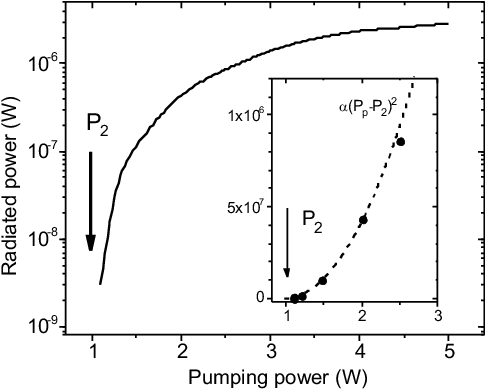}}
\end{center}
\vspace*{-0.4cm}\caption{The power of the detected radiation at $2f_\mathrm{min}$ versus the pumping
power (note the logarithmic scale). Inset: the onset of the radiation shown in the linear scale,
illustrating the procedure for determination of the threshold $P_2$.} \label{fig3}
\end{figure}

Figure 3 shows the BEC radiation power dependence on the applied pumping power. When the pumping power
exceeds the critical power $P_2$ (close to 1\,W), the radiation power (at the frequency
$2f_\mathrm{min}$) increases tremendously from the value below the sensitivity of the detecting scheme
($<10^{-14}$\,W) to the value of about $10^{-9}$\,W. The critical power $P_2$ was determined by the
extrapolation of the intensity of the detected radiation to zero, as shown in the inset of
Fig.~\ref{fig3}.  The radiation power at double frequency from a magnetic medium is proportional to the
second time derivative of the magnetization of the medium squared
$\propto(\mathrm{d}^2\mathrm{\textbf{M}}/\mathrm{d}t^2)^2$. In the case of condensate the derivative is
proportional to the density of the condensate magnons $\mathrm{d}^2 \mathrm{\textbf{M}} / \mathrm{d} t^2
\propto N_\mathrm{\textbf{k}_\mathrm{min}}$ \cite{ref10}. As the result, the radiation power at
$2f_\mathrm{min}$ is proportional to $N_\mathrm{\textbf{k}_\mathrm{min}}^2$, or close to the threshold
to $(P_\mathrm{p}-P_2)^2$. As seen from Fig.~\ref{fig3}, the experimental data nicely corroborate this
theoretical prediction. For higher pumping powers the measured function deviates from the quadratic
dependence with the tendency to saturate.

\begin{figure}[b]
\begin{center}
\scalebox{1}{\includegraphics[width=8.0 cm,clip]{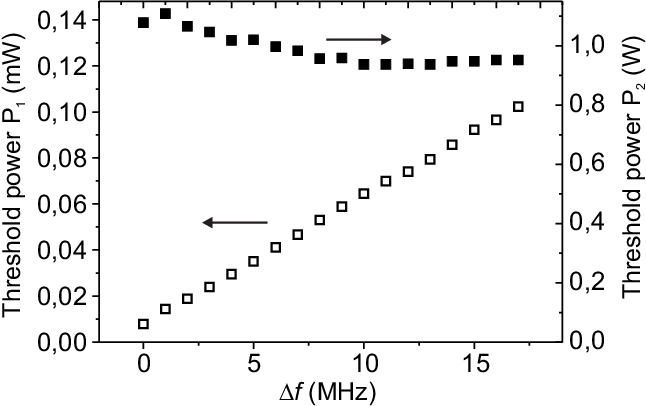}}
\end{center}
\vspace*{-0.4cm}\caption{Dependencies of the two thresholds $P_1$ ($\square$) and P2 ($\blacksquare$) on
the spectral width $\Delta f$. Note the linear dependence for $P_1$ and the almost constant value for
$P_2$.} \label{fig4}
\end{figure}

Figure~\ref{fig4} demonstrates the dependence of the critical powers $P_1$ and $P_2$ on $\Delta f$. One
can see that the threshold $P_1$ of the incoherent parametric pumping of magnons increases linearly with
$\Delta f$ in accordance with the theory of incoherent noise pumping \cite{ref15, ref16, ref17}. On the
contrary, the critical power $P_2$ of the BEC formation does not depend essentially on $\Delta f$,
especially for $\Delta f\geq10$\,MHz, where $P_2$ is nearly constant. This result can be explained using
kinetic equation for magnons \cite{ref10}. The balance between the excited and relaxing magnons at the
point of the BEC transition can be represented as

\begin{equation}\label{eq1}
\frac{(\gamma h_\mathrm{c})^2}{\Delta f} \frac{k_\mathrm{B} T}{8\pi^4\hbar} \int\limits_{\Delta
f}\frac{|V_\mathrm{\textbf{k}}|^2\mathrm{d}^3k}{f_\mathrm{\textbf{k}}-f_\mathrm{min}}\simeq\frac{1}{\tau_\mathrm{N}}
[N(\mu_\mathrm{c},T)-N(0,T)],
\end{equation}
where $\gamma$ is the gyromagnetic ratio, $h_\mathrm{c}$ is the critical amplitude of the pumping field
$(P_2\propto h_\mathrm{c}^2)$ , $V_\mathrm{\textbf{k}}$ is the magnon pair coupling strength with the
microwave field, $\tau_\mathrm{N}$ is the relaxation time of magnons, and $N(\mu_\mathrm{c},T)$ and
$N(0,T)$ are the volume densities of magnons at the point of the BEC transition and at the true
equilibrium with the lattice, correspondingly. The left side of Eq.~(\ref{eq1}) describes the density of
parametric magnons excited in the frequency interval $f_\mathrm{p}\pm\frac{1}{2}\Delta f$ per unit time.
Since for $\Delta f/f_\mathrm{p}\ll1$ $V_\mathrm{\textbf{k}}$ can be considered as a constant, the
integral in Eq.~(\ref{eq1}) is just proportional to $\Delta f$. Therefore, the left side is
approximately independent of $\Delta f$ and can be expressed as $const\cdot h_\mathrm{c}^2\propto P_2$.
Since the right side of Eq.~(\ref{eq1}), which is the relaxation term, does not depend on the external
pumping at all, one gets that $P_2$ is independent of $\Delta f$.

A slight increase of $P_2$ with the lowering of $\Delta f$ below 10\,MHz (see Fig.~\ref{fig4}) can
indicate the growing role of phase relations between the pump field and the parametric pairs, which is
not taken into account in Eq.~(\ref{eq1}). The mechanism of phase mismatching restricts the nonlinear
spin-wave excitation and therefore decreases the number of the excited magnons at a given power of the
applied microwave field.

Note that the BEC of magnons occurs in our experiment at rather large supercriticalities $P_2/P_1\gg1$.
Our current setup does not allow increasing $\Delta f$ above 17\,MHz. However, extrapolating the data in
Fig.~\ref{fig4} to larger $\Delta f$, we can assume that $P_1$ and $P_2$ will be close to each other at
$\Delta f\simeq 200$\,MHz and the BEC transition can be achieved without strong ``overheating'' of the
parametrically excited magnons. This assumption is in qualitative agreement with the theory of BEC of
magnons under incoherent pumping \cite{ref10}. The theory predicts a possibility of BEC even in the case
when the spectrum of the incoherent pumping is so wide that threshold $P_1$ is never achieved. In this
case the magnon system will resemble the famous Frohlich's model \cite{ref2} predicting long-range
coherence and energy storage to the lowest level in the living systems.

In conclusion, we have experimentally investigated Bose-Einstein condensation of magnons pumped by an
incoherent pumping with a variable spectral width $\Delta f$. We show experimentally that the BEC
formation in a quasi-equilibrium gas can be achieved under such conditions and the power radiated by the
condensate scales as $(P_\mathrm{p}-P_2)^2$  above the BEC threshold $P_2$ in agreement with the theory.
We demonstrate that while the threshold of incoherent parametric pumping grows proportionally to $\Delta
f$, $P_2$ is essentially independent on $\Delta f$.

Support by the Deutsche Forschungsgemeinschaft (DE 639/5 and SFB/TRR 49) and by the Fundamental
Researches State Fund of Ukraine is gratefully acknowledged.

\end{document}